\documentclass[12pt]{article}
\usepackage{amsfonts,amsmath,epsfig,graphicx,amssymb}
\usepackage{amsmath,epsfig,graphicx,amssymb}
\usepackage[margin=1in]{geometry}
\usepackage{amsmath}
\usepackage{amsfonts}
\usepackage{amssymb}
\usepackage{makeidx}
\usepackage{graphicx}
\newcommand{\beq}{\begin{equation}}
\newcommand{\eeq}{\end{equation}}
\newcommand{\ben}{\begin{eqnarray}}
\newcommand{\een}{\end{eqnarray}}
\date{}
\begin{document}
\title{A Conformally Invariant Unified Theory of Maxwell Fields and Linearized Gravity as Emergent Fields}
\author{Partha Nandi\footnote{parthanandyphysics@gmail.com, parthanandi@bose.res.in}\\S. N. Bose National Centre for Basic Sciences,\\Block JD, Sector III, Bidhannagar, Kolkata 700106\\and\\
Partha Ghose\footnote{partha.ghose@gmail.com} \\
Tagore Centre for Natural Sciences and Philosophy,\\ Rabindra Tirtha, New Town, Kolkata 700156, India}
\maketitle
\begin{abstract}
A Lorentz and conformally invariant `Schr\"{o}dinger-like' equation for a massless complex scalar function $\psi$ is derived from an invariant action, and it is shown how the same $\psi$ can be used to calculate both the gravitational field $h_{\mu\nu}$ of linearized Einstein gravity in the TT gauge and the electromagnetic field $F_{\mu\nu}$ in the Lorenz gauge, and that the main difference between classical and quantum aspects of such fields lies in a certain condition that the underlying scalar wave must satisfy to keep it nondispersive. It is also shown how the existence of gravitons can be inferred from the state dependent quantum noise they create in a model detector. 
\end{abstract}

\section{Introduction}
Although there is no empirical evidence that gravity is quantized, quantum gravity has been the holy grail of theoretical physics for nearly a century because of the enormous success of both Einstein's general relativistic theory of gravity and quantum mechanics. Among the many predictions of Einstein's theory of gravity that have passed the test, the detection of gravitational waves by the LIGO-Virgo collaboration is the latest feather in the cap \cite{ligo}. These gravitational waves are enormously weak, and it is quite legitimate to use the linearized version of Einstein gravity for them. The question is: is there any empirical way to find out if these linearized waves are quantized? Recently, Parikh, Wilczek and Zahariade \cite{par,ef,FZ} have shown that if gravitons exist, they will produce characteristic quantum noise in the detector, signalling their presence.\\

Another holy grail of physics has been the unification of gravity and electrodynamics. Einstein's goal was more ambitious, namely to construct a unitary theory that would account for both unification and quantization. Early attempts at unification turned out to be unsatisfactory as they did not take into account the quantum nature of radiation \cite{goenner}. \\

In this paper we propose the fundamentals of a single relativistic complex scalar wave theory that underlies both {\em linearized} Einstein gravity and electrodynamics and accounts for their unification and quantization. That a complex scalar representation $\psi$ of the free electromagnetic field exists was shown by Wolf and Green in the 1950s \cite{green, wolf}. Their motivation, however, was limited to the practical ease of calculating the vector potential and free classical fields from a scalar $\psi$. From our point of view, the crucial point is the use of a {\em complex} $\psi$ which is the hallmark of quantum mechanics. The missing links in their work were the precise mathematical relations between $\psi$ and the gauge potentials (and hence fields) in electrodynamics and linearized gravity, and also the Schr\"{o}dinger-like dynamical equation that such a $\psi$ satisfies. We will specify these relations and derive the Schr\"{o}dinger-like equation for $\psi$ from the second order classical wave equation that it satisfies, revealing the underlying unity of classical and quantum {\em wave mechanics}. It turns out that the difference between the two lies not in the dynamical equations but in a certain condition that only a classical wave amplitude satisfies that renders it non-dispersive. This is different from the standard expectations concerning the relationship between the nonrelativistic classical and quantum mechanics of {\em particles}.\\

The plan of the paper is as follows. In Section 2, we demonstrate the equivalence of a {\em complex} classical gauge theory to linearized gravity in the TT gauge. This is different from the standard versions of gravitoelectrodynamics which use a real gauge field \cite{mash}. In section 3 we show that this complex gauge theory can be constructed from an underlying complex scalar wave theory which is classical or quantum mechanical depending on whether or not the wave amplitude satisfies a certain condition. It will also be shown how to isolate the pure quantum effects from the classical ones by using the Hamilton-Jacobi method.  The quantum effects turn out to be state dependent. in Section 4 we show how this purely state dependent quantum noise modifies the detector response. Finally in section 5 we show how the same scalar wave function $\psi$ is related to the electromagnetic vector potential and fields.

\section{Equivalence of A Complex Gauge Theory and Linearized Einstein Gravity}

The Einstein field equations of General Relativity,
\beq
R_{\mu\nu} - \frac{1}{2}g_{\mu\nu}R = \kappa T_{\mu\nu},
\eeq
follow as variational equations from the Einstein-Hilbert action
\beq
S = \int \left[\frac{1}{2\kappa}R + {\cal L}_{M}\right]\sqrt{-g} d^4x
\eeq
where $R$ is the Ricci scalar, ${\cal L}_{M}$ is the matter Lagrangian and $\kappa = 8\pi G/c^4$.
In the weak field approximation,
\beq
g_{\mu\nu} = \eta_{\mu\nu} + h_{\mu\nu},\,\,\,\,|h_{\mu\nu}|\ll 1,
\eeq
one has
\beq
R =\eta^{\mu\nu}R_{\mu\nu} = \partial_\mu\partial_\nu h^{\mu\nu} - \Box h \label{R}
\eeq
where $h = \eta^{\mu\nu}h_{\mu\nu}$ is the trace of $h_{\mu\nu}$. It is useful to define the trace-reversed amplitude $\bar{h}_{\mu\nu} = h_{\mu\nu} - \frac{1}{2}\eta_{\mu\nu}h$, and to linear order in the perturbation and in the gauge $\bar{h}^{\mu\nu}_{\,\,\,\,,\nu} = 0$, one gets the equations
\beq
\Box\bar{h}_{\mu\nu} = \kappa T_{\mu\nu}.
\eeq
Henceforth, we will restrict ourselves to the solutions of the homogeneous equations
\beq
\Box h_{\mu\nu} = 0
\eeq
derivable from the action
\beq
S_g = \int\frac{1}{2\kappa}R\sqrt{-g} d^4x,\,\,\,\,R = -\Box h.
\eeq
We will see that in this linearized version Einstein gravity reduces to gravitoelectrodynamics (GEM) paralleling Maxwell electrodynamics but with helicity $\pm 2$ \cite{mash}.

The gravitational field, being a symmetric second rank tensor, has 10 degrees of freedom. 
Let us consider the action 
\ben
S_F &=& \int {\cal L}_F d^4 x = -\frac{1}{2}\int F^{\mu\nu*}F_{\mu\nu} d^4x = \int \left(\vec{{\cal E}^*}.{\vec{\cal E}} - \vec{{\cal B}}^*.\vec{{\cal B}}\right) d^4x \label{Sp} 
\een
where $F^{\mu\nu}$ is a complex second rank antisymmetric tensor with gravitoelectric and gravitomagnetic components $\vec{{\cal E}} = \frac{1}{\sqrt{2}} (\vec{E}_1 + i\vec{E}_2)$ and $\vec{{\cal B}} = \frac{1}{\sqrt{2}}(\vec{B}_1 + i\vec{B}_2)$ respectively, with the transversality conditions $\vec{E}_1.\vec{B}_1 = \vec{E}_2.\vec{B}_2 =0$ which bring down the number of independent components from 12 to 10. Variations of the action lead to the Maxwell-like equations
\beq
\partial_\mu F^{\mu\nu} = 0,\,\,\,\,\partial_\mu F^{\mu\nu*} = 0.
\eeq
In addition, one has the Bianchi identities
\ben
\partial_\lambda F_{\mu\nu} + \partial_\mu F_{\nu\lambda} + \partial_\nu F_{\lambda\mu} = 0,\\
\partial_\lambda F^*_{\mu\nu} + \partial_\mu F^*_{\nu\lambda} + \partial_\nu F^*_{\lambda\mu} = 0.
\een
Defining 
\ben
F_{\mu\nu} &=& \partial_\mu A_\nu - \partial_\nu A_\mu,\\
h^{\mu\nu} &=& \kappa (A^{\mu *} A^\nu + A^{\nu *}A^\mu), \label{h}
\een
and using $\eta_{\mu\nu}$ and $\eta^{\mu\nu}$ to lower and raise indices, we have (see Appendix)
\ben
{\cal L}_F &=& - \frac{1}{2}F^{\mu\nu*}F_{\mu\nu} = \partial^\mu A^{\nu*}\partial_\mu A_\nu - \partial^\mu A^{\nu *}\partial_\nu A_\mu = - \left[\Box A_\mu - \partial_\nu \partial_\mu A^\nu   \right] A^{\mu *}\nonumber\\
&=& -\frac{1}{2\kappa}\Box h + A_\mu \Box A^{\mu ^*} - 2 \partial_\lambda (A_\mu \partial^\lambda A^{\mu ^*})  
\een
The last term is a total 4-divergence and can be dropped without affecting the equations of motion. Hence, the action $S_F$  of a complex gauge theory (eqn (\ref{Sp})) with the transversality conditions $\vec{E}_1.\vec{B}_1 = \vec{E}_2.\vec{B}_2 =0$ is equivalent to that of linearized Einstein gravity in the transverse-traceless (TT) gauge in which the term $\partial_\mu \partial_\nu h^{\mu\nu} = 0$ corresponding to the Lorenz gauge in which $\Box A^{\mu ^*} = 0$.

\section{Quantum Mechanics of Linearized Gravity}

Green and Wolf \cite{green, wolf} have shown that electromagnetic fields in vacuum can be rigorously derived from a complex scalar field. Following a similar approach for complex gravitoelectrodynamics, one can write
\ben
S_\psi = \int {\cal L}_\psi d^4x &=& \xi\int \left[\partial^\mu \psi^*(x) \partial_\mu\psi(x)\right]d^4x \label{L1}
\een
where the constant $\xi =\hbar l_P$ with $l_P$ as the Planck length is introduced to make $\psi^*\psi$ have the dimension of $L^{-3}$. The variational equation that follows is
\beq
\Box \psi(x) = 0. \label{L2}
\eeq 
Let
\beq
\psi(x) = \int \frac{d^3k}{(2\pi)^32\omega(k)}\phi(k)\exp (-ikx)
\eeq
be a solution. Given $\psi$, one can compute the gravitoelectromagnetic gauge potentials $A_\mu$ (and hence $h_{\mu\nu}$, using definition (\ref{h})) as follows.
Define $A_\mu$ as
\beq
A_\mu = \sqrt{\xi}\, \int \frac{d^3k}{(2\pi)^32\omega(k)}\sum_{\lambda=1,2}\epsilon^{(\lambda)}_\mu(k)\phi(k)\exp (-ikx)\label{a}
\eeq
where $\epsilon^{(\lambda)}_\mu(k)$ is a unit polarization vector (with $\lambda = 1,2$ to include only the physical transverse degrees of freedom) satisfying the orthogonality condition
\beq
\epsilon^{\mu(\lambda)*}(k)\epsilon^{(\lambda^\prime)}_\mu(k) = \delta^{\lambda\lambda^\prime}.
\eeq 
Hence, $\Box A_\mu = 0$. One can then define the gauge invariant field $F_{\mu\nu} = \partial_\mu A_\nu - \partial_\nu A_\mu$ and derive the equation $\partial^\mu A_\mu = 0$ from the equation $\partial^\mu F_{\mu\nu} = 0$ on the assumption that the potential $A_\mu$ vanishes at infinity. This is the Lorenz gauge.
This completely specifies the representation of linearized gravity in terms of a complex scalar wave function, and shows that the Planck length $l_P$ enters such a theory of gravity through the action $S_\psi$.

The variational equation (\ref{L2}) is Lorentz and conformal invariant. Since an arbitrary wave packet can be constructed from monochromatic plane waves in classical physics, consider the monochromatic solutions \cite{Wqm}
\ben
\psi(x) &=& A_k e^{-ikx} = A_k e^{i(\vec{k}.\vec{x} - k_0x_0)},\,\,\,\,k_0 = \frac{\omega}{c}.
\een
We get
\ben
(\partial^{\prime 2}_0 - \nabla^2_{\vec{x}^\prime}) \psi^\prime(x^\prime) &=& (-ik_0^\prime\partial^\prime_0 -\nabla^2_{\vec{x}^\prime})\psi^\prime(x^\prime)\\
&=&  (\partial_0^2 - \nabla^2_{\vec{x}})\psi(x)\\
&=& (-ik_0\partial_0 -\nabla^2_{\vec{x}})\psi(x) = 0.
\een
Thus, one obtains the Lorentz and conformal invariant equation
\beq
i\dot{\psi} = - \frac{c}{k_0}\nabla^2\psi. \label{Q}
\eeq
Comparing with the nonrelativistic Schr\"{o}dinger equation 
\beq
i\dot{\psi} = - \frac{\hbar}{2m}\nabla^2 \psi
\eeq
one finds a surprising `correspondence': the left hand sides are identical, and the right hand sides differ only in the coefficient of $\nabla^2$. However, note that 
\beq
\frac{c}{k_0} = \frac{\hbar}{\hbar k_0/c} := \frac{\hbar}{2m^*},\,\,2m^* = \frac{\hbar k_0}{c}
\eeq 
and hence eqn (\ref{Q}) can be written in the form
\beq
i\dot{\psi} = - \frac{\hbar}{2m^*}\nabla^2\psi \label{sch}
\eeq
with $m^*$ as an `effective mass' which transforms like $k_0$ under Lorentz transformations. This is therefore a relativistic and conformally invariant `Schr\"{o}dinger-like' equation for a massless particle, and can act as the basis of the quantum mechanics of radiation. The transition from a classical wave function to a quantum-like wave function is facilitated by the introduction of the Planck constant which introduces a new scale through the relation $E = \hbar\omega$ which, together with the Einstein relation $E = mc^2$, implies $m = \hbar\omega/c^2$, i.e. that mass and frequency are equivalent. This holds for every monochromatic component of a wave packet.

If one applies the remaining time derivative in (\ref{Q}) on $\psi$ and use the definition $k_0 c = \omega$, one gets the classical Helmholtz equation
\beq
\left(\nabla^2 + k^2\right)\psi = 0,\,\,\,\,k^2 = \frac{\omega^2}{c^2} \label{Helm}
\eeq
where $k$ is the wave number in vacuum (refractive index $n = 1$).
Although this equation is derivable from eqn (\ref{Q}), there is a crucial difference between them. And, although eqn (\ref{Q}) was derived using a monochromatic solution of the wave equation (\ref{L2}), it admits of more general solutions. Consider a general solution of in the polar form  
\ben
\psi(x) &=& \sqrt{\rho(x)}\,{\rm exp}(i\phi(x)),\label{polar1}\\
\phi(x) &=& \vec{k}.\vec{x} - \omega t \label{polar2}
\een
where $\rho(x)$ and $\phi(x)$ are real Lorentz scalar functions. Separating the real and imaginary parts, one obtains  
\beq
k^2 = \frac{\omega^2}{c^2} + \frac{\nabla^2 \sqrt{\rho (x)}}{\sqrt{\rho (x)}}. \label{disp2}
\eeq
However, substituting the same solutions for $\psi$ in the classical equation (\ref{Helm}) and separating the real and imaginary parts result in the constraint
\beq
\frac{\nabla^2 \sqrt{\rho (x)}}{\sqrt{\rho (x)}} = 0\label{qp}
\eeq
from the real part, which is consistent with eqn (\ref{Helm}). The additional $x$-dependent term in eqn (\ref{disp2}) causes dispersion but vanishes in the classical case, ensuring that classical wave packets are non-dispersive. Hence, wave functions that satisfy eqn (\ref{Q}) (and therefore (\ref{sch})) are {\em non-classical} in the sense that, unlike classical wave functions, they do not satisfy condition (\ref{qp}). 

These non-classical solutions turn into quantum mechanical wave functions on introducing a quantized scale of action $\hbar$. Then the wave functions in the action $S_\psi$ cannot be arbitrarily scaled and become normalized:
\beq
\int \psi^*\psi\, d^3x = 1. \label{norm}
\eeq 
The convection current 
\beq
\vec{j} = -\frac{i\hbar}{2}\left[\psi^* \vec{\nabla}\psi -\vec{\nabla}\psi^* \psi\right]\label{cur}
\eeq
associated with the Schr\"{o}dinger-like equation (\ref{sch}) satisfies the continuity equation
\beq
\partial_\mu j^\mu = \frac{\partial \rho}{\partial t} + \vec{\nabla}.\vec{j} = 0
\eeq
with $j_0 =\rho =\hbar k_0|\psi|^2 >0$, which enables a probabilistic interpretation of $\psi$ (the Born rule) satisfying the normalization condition (\ref{norm}) \cite{note1}.
 
Writing $\phi = S/\hbar$, one can write the general solution (\ref{polar1}) in the form
\begin{eqnarray}
\psi(x) &=& \sqrt{\rho(x)}\,{\rm exp}(iS(x)/\hbar), \label{polar}
\end{eqnarray}
where both $\rho$ and $S$ are real Lorentz scalar functions. Separating the real and imaginary parts and substituting in eqn.(\ref{sch}), one obtains the coupled equations
\ben
\frac{\partial S(x)}{\partial t} + \frac{(\nabla S(x))^2}{2m^*} + Q &:=& \frac{\partial S(x)}{\partial t} + H = 0, \label{Hamq}\\
Q &=& - \frac{\hbar^2}{2m^*}\,\frac{\nabla^2 \sqrt{\rho(x)}}{\sqrt{\rho(x)}},\label{Q1}
\een
and
\begin{eqnarray}
\frac{\partial \rho}{\partial t} + \vec{\nabla}.(\rho \vec{\nabla} W) = 0.\label{conteq}
\end{eqnarray}
These are the conditions that must be satisfied for general solutions of eqn (\ref{sch}).
Eqn. (\ref{Hamq}) is the Hamilton-Jacobi equation in gravitoelectrodynamics and Eqn. (\ref{conteq}) is a conservation law (the gravitational Poynting theorem). $Q$ is known in the literature as the `quantum potential'. This shows that $Q$ is the {\em sole} correction to the classical Hamilton-Jacobi equation, vanishing as it does by condition (\ref{qp}) in classical wave theory, {\em independent of $\hbar$}. One knows that $\hbar \neq 0$ in the actual world which nevertheless has an undeniable classical aspect which is guaranteed by condition (\ref{qp}) on the wave amplitude. Hence, $Q$ neatly isolates the purely non-classical and therefore quantum `noise' in a quantum mechanical wave.

It is important to note that the polar representation (\ref{polar}) of the wave function is a purely mathematical transformation without necessarily implying the well known de Broglie-Bohm theory which also uses the same transformation but is distinguished from standard quantum mechanics by an additional hypothesis, namely the guidance condition $\vec{p} = \vec{\nabla}S$ which we have not used.

Given the quantum mechanical interpretation of $\psi$ satisfying eqn (\ref{sch}), we have a quantum mechanics of linearized gravity using the relations (\ref{a}) and (\ref{h}). 
 
\section{Detection of Gravitons}
Finally, let us consider the detection of gravitational waves. For simplicity let us assume that the detector is a classical harmonic oscillator of mass $m$ whose Hamiltonian is
\beq
H_0^d = \frac{1}{2m}\vec{p}.\vec{p} + \frac{1}{2}m\omega^2 \vec{x}.\vec{x}
\eeq 
Let it interact with the gravitational wave at $x = (t,\vec{x})$, and let the interaction be of the minimal type. Then, following DeWitt \cite{dW}, we can write
\beq
H^d = \frac{1}{2m}\left(\vec{p} - \vec{\mathcal{B}}\right)^2 + \frac{1}{2}m\omega^2 \vec{x}.\vec{x} + V
\eeq
where
\ben
V &=& - \frac{1}{2}mc^2 h_{00} = -mc^2 \kappa A_0^*A_0,\\
\mathcal{B}_i &=& mc h_{oi} = mc\kappa \left(A_0 A_i^* + A_0^* A_i\right). 
\een

The total Hamiltonian of the system (detector + gravitational wave) is thus
\ben
H &=& H^d + H^g,\\
H^g &=& H_0^g + Q
\een
where $H_0^g$ is the classical Hamiltonian of the gravitational field and $Q$ is the purely quantum contribution to it, given by eqn (\ref{Q1}). Hence, the equation of motion of the detector coordinate is
\beq
\dot{p}_i = - \frac{\partial H}{\partial x_i} = - \frac{\partial (H^d + H_0^g)}{\partial x_i} - \frac{\partial Q}{\partial x_i}. 
\eeq
The last term is the contribution of the pure quantum mechanical part of the gravitational wave (the quantum noise). It is clear from eqn (\ref{Q1}) that this force is dependent on the functional nature of the wave amplitude $\sqrt{\rho}$ of the gravitational wave. This is a very simple and intuitive demonstration of the essential result emphasized in Parikh, Wilczek and Zahariade \cite{par}.

{\flushleft{{\em Gravitational Coherent State}}}

Let us now look at a particular state that is closest to the classical case, namely the coherent state of gravitons. Consider the gravitational Schr\"{o}dinger equation
\begin{equation}
	i\hbar\frac{\partial}{\partial t}\Psi(\vec{r},t)=-\frac{\hbar^{2}}{2m^*}\nabla^{2} \Psi(\vec{r},t),
\end{equation}
with $m^*=\frac{\hbar \omega}{2c^{2}}$.
It is useful to introduce the dimensionless variables
\begin{equation}
\vec{q}=l^{-1}_{P}\vec{r},~~~~~ \tau=\frac{\hbar}{m^*l^{2}_{p}}t,
\end{equation}
where $l_{P}=\sqrt{\frac{\hbar G}{c^{3}}}$.
Therefore, the above gravitational Schr\"{o}dinger equation becomes
\begin{equation}
i\frac{\partial}{\partial \tau}\Psi(\vec{q}.\tau)=-\frac{1}{2}\frac{\partial^{2}}{\partial q^{2}}\Psi(\vec{q},\tau)
\end{equation} 
with $[\hat{q}_{i},\hat{p}_{j}]=i\delta_{ij}$ where the coordinate representation of the spatial translation generator:  $\hat{p}_{j}=-i\frac{\partial}{\partial q_{j}}.$\\

Let us introduce step-up (and corresponding step-down) operators  
\begin{equation}
\hat{a}_{i}=\frac{\hat{q}_{j}+i\hat{p}_{j}}{\sqrt{2}},
\end{equation}
satisfying the commutation relation 
\begin{equation}
[\hat{a}_{i},\hat{a}^{\dagger}_{j}]=\delta_{ij},
\end{equation}
where $i= 1, 2, 3$.
 The bosonic realization of the Heisenberg Weyl Lie algebra will be furnished by the underlying Hilbert space which is isomorphic to the bosonic Fock space:
\begin{equation}
 \mathcal{H}= {\rm span}\{\left|n_{1},n_{2},n_{3}\right\rangle= \frac{1}{\sqrt{n_{1}!~n_{2}!~n_{3}!}}(\hat{a_{1}}^{\dagger})^{n_{1}}~(\hat{a}_{2}^{\dagger})^{n_{2}}~(\hat{a}_{3}^{\dagger})^{n_{3}}\left|0,0,0\right\rangle\}
 \label{hc}
\end{equation}
Following Glauber, the coherent state may be defined
through a superposition of the occupation number states $\left|n_{1},n_{2},n_{3}\right\rangle,$ by requiring that it be a simultaneous eigenstate of
the commuting annihilation operators $\hat{a}_{1},\hat{a}_{2},\hat{a}_{3}$ with the eigenvalues $\alpha_{1},\alpha_{2}, \alpha_{2}$ \cite{BDR}:
\begin{equation}
\left|\alpha_{1},\alpha_{2},\alpha_{3}\right\rangle=e^{-\frac{1}{2}(\mid\alpha_{1}\mid^{2}+\mid\alpha_{2}\mid^{2}+\mid\alpha_{3}\mid^{2})}\sum_{n_{1},n_{2},n_{3}=0}^{n_{1},n_{2},n_{3}=\infty} \frac{\alpha^{n_{1}}_{1}~\alpha^{n_{2}}_{2}~\alpha^{n_{3}}_{3}}{\sqrt{n_{1}!~n_{2}!~n_{3}!}}~\left|n_{1},n_{2},n_{3}\right\rangle
\end{equation}
These states provide an over-complete basis on $\mathcal{H}$ and it is possible to write the identity as
\begin{equation}
\int\prod_{i=1}^{3} \frac{d\alpha_{i} d\bar{\alpha}_{i}}{\pi^{3}}  \left|\alpha_{1},\alpha_{2},\alpha_{3} \right\rangle \langle \alpha_{1},\alpha_{2},\alpha_{3}|=\mathbb{I},
\end{equation}
Furthermore,  the nature of the coherent state in the co-ordinate representation,
\begin{equation}
\langle q_{1}~q_{2}~q_{3}\left|\alpha_{1},\alpha_{2},\alpha_{3}\right\rangle=\left(\frac{1}{\pi}\right)^{\frac{3}{4}}~e^{-\frac{1}{2}\sum_{i=1}^{3}(\mid\alpha_{i}\mid^{2} +q^{2}_{i})}\sum_{n_{1},n_{2},n_{3}=0}^{\infty} \prod_{j=1}^{3}\frac{\alpha^{~n_{j}}_{j}}{\sqrt{n_{j}~!}}\frac{i^{n_{j}}}{\sqrt{2^{n_{j}} n_{j}!}} H_{n_{j}}(q_{j}),
\end{equation}
where $H_{n_{j}}(q_{j})$is the the Hermite polynomial of order $n_{j}.$\\

Now, making use of the generating function for
Hermite polynomials
\begin{equation}
e^{2tq_{j}-q^{2}_{j}}=\sum_{n=0} ^{\infty}\frac{H_{n}(q_{j}) ~t^{n}}{n!},
\end{equation}
we can immediately throw the co-ordinate state representation
of the coherent state into the form
\begin{equation}
\Psi(\vec{q}~;~\{\alpha_{i}\})=\langle q_{1}~q_{2}~q_{3}\left|\alpha_{1},\alpha_{2},\alpha_{3}\right\rangle=(\frac{1}{\pi})^{\frac{3}{4}}e^{-\frac{1}{2}\sum_{i=1}^{3}(\mid\alpha_{i}\mid^{2}+\alpha^{2}_{i})} \prod_{j=1}^{3} e^{-\frac{1}{2}(q_{j}-i\sqrt{2}\alpha_{j})^{2}}
\end{equation}
Let us define 
\begin{equation}
 \alpha_{i}=Re(\alpha_{i})+i Im(\alpha_{i})
 \end{equation}
 
 \begin{equation}
 \Psi(\vec{q}~;~\{\alpha_{i}\})\approx \sqrt{\rho(\vec{q;\{\alpha_{i}\}})}~e^{i\frac{S(\vec{q},\{\alpha_{i}\})}{\hbar}}\label{psi}
 \end{equation}
where
\begin{equation}
 S=-\hbar\sum_{i=1}^{3}[Re(\alpha_{i})Im(\alpha_{i})-\sqrt{2}(q+\sqrt{2}Im (\alpha_{i}))]
 \end{equation}
 
\begin{equation}
  \rho(\vec{q};\{\alpha_{i}\})= \prod_{i=1}^{3} e^{-(q_{i}+\sqrt{2}Im(\alpha_{i}))^{2}}
  \label{G}
  \end{equation}
 As $\vec{q}=l^{-1}_{P}\vec{r}$  we may write 
 
\begin{equation}
 \rho(\vec{	r})= \prod_{i=1}^{3} e^{-\frac{1}{l_{P}^{2}}[(r_{i}+\sqrt{2} l_{P}Im(\alpha_{i}))^{2}}
\end{equation}
This is a probability distribution due to a displacement of the ground state wave-function for a three-dimensional oscillator, representing a displaced Gaussian pattern. 

Thus, the quantum potential turns out to be
\begin{equation}
 Q=\frac{\hbar^{2}}{2m_{\omega}\sqrt{\rho(\vec{r})}}~\vec{\nabla}^{2}~\sqrt{\rho(\vec{r})}=\frac{\hbar^{2}}{2m_{\omega}l_{P}^{4}}[\sum_{i=1}^{3}(r_{i}+\sqrt{2} l_{P}Im(\alpha_{i}))^{2}-3l^{2}_{P}]
 \end{equation}
This is the quantum noise in the gravitational wave. Its detection is tantamount to a measurement of the quantum potential. Note that it is singular in $\hbar$ like the Aharonov-Bohm phase and therefore it might appear that there is no classical limit. This is not true because of condition (\ref{qp}) which guarantees classicality even in the presence of a non-vanishing $\hbar$. 

The displaced Gaussian pattern of the graviton's wave function (a coherent state) will clearly show up in the intensity pattern of an interference experiment \cite{baxter, Ba} involving the superposition of two identical coherent state wave packets (\ref{psi}).

{\flushleft{\em Quantum vs Classical Interference}}

Let us consider a double-slit set up and write a coherent superposition of two identical
displaced Gaussian wave packets at some point on a screen with radius vector $\vec{r}$:

\begin{equation}
\psi(\vec{r}~;~\{\alpha_{i}\}) = \frac{1}{\sqrt{2}}\left[\sqrt{\rho(\vec{r} - \vec{r}_1)}~e^{i\frac{S_{1}(\vec{r} - \vec{r}_1)}{\hbar}} + \sqrt{\rho(\vec{r} - \vec{r}_2))}~e^{i\frac{S_{2}(\vec{r} - \vec{r}_2)}{\hbar}}\right]
\end{equation}
where $\vec{r}_1$ and $\vec{r}_2$ are the radius vectors of the two slits. When both the slits are open, an interference pattern will occur with the intensity of the sinusoidal term given by 
\begin{equation} 
\sqrt{\rho(\vec{r}-\vec{r}_{1})\rho(\vec{r}-\vec{r}_{2})}=e^{-\frac{[\mid \vec{r}-\vec{r}_{1}\mid^{2}+\mid \vec{r}-\vec{r}_{2}\mid^{2}]}{2l_{P}^{2}}} e^{-\frac{1}{l_{P}}[\sum_{i}\sqrt{2}Im(\alpha_{i})(2\vec{r}_{i}-(\vec{r}_{1}+\vec{r}_{2})_{i})+2l_{P}\sum_{i}Im^{2}(\alpha_{i})]},
\end{equation} 
 This carries a distinct quantum signature that would be absent in classical wave packets.
 
\section{Quantum Mechanics of Electromagnetic Radiation}

It turns out that the same action $S_\psi$ (\ref{L1}) also describes free electromagnetic radiation in vacuum provided the real electromagnetic gauge potentials ${\cal A}_\mu$ are given by the real part of the complex gauge potentials $A_\mu$ (eqn (\ref{a})).

This specifies the correspondence completely. Thus, the complex wave function $\psi$ underlies linearized Einstein gravity while its real part underlies electromagnetic radiation. 

\section{Concluding Remarks}
Starting from an invariant action, we have derived a relativistic and conformally invariant `Schr\"{o}dinger-like' equation for a massless complex scalar wave function $\psi$ which can be used to calculate both the gravitational field $h_{\mu\nu}$ of linearized Einstein gravity in the TT gauge and the electromagnetic field $F_{\mu\nu}$. This shows that both unification and quantization can be achieved at the level of a single underlying complex scalar function which is common to both these fields which turn out to be emergent rather than fundamental fields. This is significant from the methodological point of view.

From the technical point of view the unification required the generalization of standard gravitoelectromagnetism to a theory with a {\em complex} potential $A_\mu$ and a complex field $F_{\mu\nu}$ satisfying certain orthogonality conditions. 

The Hamilton-Jacobi method proved useful to isolate the purely quantum contributions of the fields to the Hamiltonian. A simple model is given to show how the purely quantum noise of gravitons can be detected, and how it is state dependent as pointed out in Ref. \cite{par}.  

An important aspect of the conformally invariant theory is the inherent relationship between the classical and quantum wave functions and their dynamical equations. This is radically different from all standard approaches to quantum-classical correspondence involving massive particles. 

It is also significant that the Planck length $l_P$ enters the theory already at the low energy scales of electromagnetic radiation and linearized gravity.

\section{Appendix}
\ben
\Box (A_\mu A^{\mu *}) &=& \partial^\lambda\left(\partial_\lambda A_\mu A^{\mu *} + A_\mu \partial_\lambda  A^{\mu *}\right)\nonumber\\
&=& \Box A_\mu A^{\mu *} + 2 \partial_\lambda A_\mu \partial^\lambda A^{\mu *} + A_\mu \Box A^{\mu *}
\een
and hence
\beq
\Box A_\mu A^{\mu *} = \Box (A_\mu A^{\mu ^*}) - 2 \partial_\lambda A_\mu \partial^\lambda A^{\mu *} - A_\mu \Box A^{\mu *}
\eeq
Further,
\beq
2 \partial_\lambda A_\mu \partial^\lambda A^{\mu *} = 2 \partial_\lambda (A_\mu \partial^\lambda A^{\mu ^*}) - 2 A_\mu \Box A^{\mu ^*}
\eeq
Therefore,
\beq
\Box A_\mu A^{\mu *} = \Box (A_\mu A^{\mu ^*}) + A_\mu \Box A^{\mu ^*} - 2 \partial_\lambda (A_\mu \partial^\lambda A^{\mu ^*}) 
\eeq


\begin{thebibliography}{0}
\bibitem{note1}
Note the difference with the Klein-Gordon current whose time component is not positive definite.
\end{thebibliography}
\end{document}